\documentclass[superscriptaddress,twocolumn,showpacs,preprintnumbers,amsmath,amssymb,aps]{revtex4}

\usepackage{graphicx}% Include figure files
\usepackage{dcolumn}% Align table columns on decimal point
\usepackage{bm}% bold math
\usepackage{bbm}

\graphicspath{{./figures/}}
\newcommand{\be}{\begin{equation}}
\newcommand{\beq}{\begin{equation}}
\newcommand{\en}{\end{equation}}
\newcommand{\eeq}{\end{equation}}
\newcommand{\bea}{\begin{eqnarray}}
\newcommand{\ena}{\end{eqnarray}}
\newcommand{\hbo}{\hbox to 1 true cm {\hfill } }
\newcommand{\tr}{\hbox{tr}}

\def\eq#1{(\ref{#1})}

\begin{document}

%\preprint{ }

\title{Two-colour QCD with heavy quarks at finite densities} 
% Force line breaks with \\

\author{Kurt Langfeld} \affiliation{School of Computing \&
  Mathematics and the Mathematical Sciences Research Centre (CMS), 
Plymouth University, Plymouth, PL4 8AA, UK  }

\author{Jan M. Pawlowski} \affiliation{Institut f\"ur Theoretische
  Physik, Universit\"at Heidelberg, Philosophenweg 16, 69120
  Heidelberg, Germany} 
\affiliation{ExtreMe Matter Institute EMMI, GSI, Planckstr. 1,
  D-64291 Darmstadt, Germany}

%\date{\today
%}% It is always \today, today,

\begin{abstract}
  We extend the density-of-states approach to gauge systems (LLR
  method \cite{Langfeld:2012ah}) to QCD at finite temperature and
  density with heavy quarks. The approach features an exponential
  error suppression and yields the Polyakov loop probability
  distribution function over a range of more than hundred orders of
  magnitude. SU(2) gauge theory is considered in the confinement and
  the high-temperature phase, and at finite densities of heavy
  quarks. In the latter case, a smooth rise of the density with the
  chemical potential is observed, and no critical phenomenon
  associated with deconfinement due to finite densities is found.
\end{abstract}

\pacs{ 11.15.Ha, 12.38.Aw, 12.38.Gc }
\keywords{ Yang-Mills theory, finite density, lattice, density of states }
%Use showkeys class option if keyword
\maketitle

The phase structure of QCD is governed by two main phenomena, quark
confinement and chiral symmetry breaking. At non-vanishing chemical
potential this phase structure is unresolved yet despite the
impressive progress made in the last two decades. On the one hand this
relates to the sign-problem which hampers lattice computations at
baryo-chemical potential $\mu_B/T\gtrsim 1$. On the other hand, first
principle continuum methods at large chemical potential are hampered
by the necessity of fully taking into account the hadronic resonance
spectrum and the increasingly complicated ground state structure of
QCD at finite density.
 
In QCD with static quarks, the confinement-deconfinement transition is
a phase transition of first or second order, depending on the rank of
the gauge group. The related order parameter is the expectation value
of the Polyakov loop in the fundamental representation, $\langle
P(\vec x)\rangle $, \cite{Polyakov:1978vu,McLerran:1981pb}. It relates
to the free energy of a single quark state. For $SU(N)$ the
Polyakov loop is an order parameter for the order - disorder
transition  the centre of the gauge group. For $SU(2)$ the centre
symmetry is $Z_2$, and the transition is of second order. It is
conjectured that it belongs to the Ising universality class. 

These findings have ever since stirred the hope that an effective
theory of the Polyakov line might be based upon a rather simple action
but still has a grasp of the confinement
mechanism~\cite{Pisarski:2000eq}.  In the last two decades, this idea
has inspired numerous Polyakov loop enhanced quark-meson models. These
models contain the Polyakov line as a remanent degree of freedom from
the gluon sector, see e.g.~\cite{Meisinger:1995ih,Langfeld:1997fi,%
  Langfeld:1998yf,Fukushima:2003fw,Ratti:2005jh,Megias:2004hj,Schaefer:2007pw}
in terms of a Polyakov loop potential. More recently, it has been
shown how these Polyakov loop enhanced low energy models of QCD are
embedded in first principle QCD with the help of functional methods,
\cite{Braun:2007bx,Fister:2013bh,Braun:2009gm,Haas:2013qwp,Fischer:2013eca}. This
entails that these model potentially give access to the phase
structure of QCD and may offer explanations of heavy-ion-collision
phenomena such as the excess in the dilepton production in collisions
with heavy nuclei~\cite{Langfeld:1997fi}.

Polyakov-loop enhanced models might also provide access to the QCD
phase diagram at low temperatures and high baryon densities. Using an
expansion with respect to inverse powers of the heavy quark mass, the
QCD partition function can be systematically reduced to a theory of
Polyakov lines, interacting with (spatial) gluons, for which the
dependence on the baryon chemical potential is
known~\cite{Langfeld:1999ig}. Integrating out the spatial gluon fields
would leave us with the PLA at finite chemical potential which is, in
fact, finite density QCD in the heavy quark limit. Many attempts have
been made to guess this PLA by using symmetry
arguments~\cite{DeGrand:1983fk} or by reducing the Polyakov line
action to the simplest case of the nearest neighbour
interactions~\cite{Karsch:1985cb}. The latter so-called SU(3) spin
model has served in the past as testbed for new simulations techniques
which are designed to circumvent the notorious sign problem of dense
matter
systems~\cite{Karsch:1985cb,Aarts:2011zn,Mercado:2012ue}. Another
recent approach to the PLA at finite densities is the strong coupling
approximation~\cite{Langelage:2010yr,Fromm:2011qi}. First studies show
that scaling violations might be mild adding predictive power to this
approach. A more systematic approach to PLA has been recently has been
recently put forward: Using the
relative-weights-method~\cite{Greensite:2012dy}, a surprisingly simple
Polyakov line effective theory has been reported in momentum space at
least in the confinement phase for the SU(2) gauge
theory~\cite{Greensite:2013yd}.

The main obstacle for the calculation of the Polyakov line effective
potential by means of Monte-Carlo simulations are cancellations that
result in a poor signal-to-noise ratio. Recently a density-of-states
method on the basis of the Wang-Landau approach~\cite{Wang:2001ab} has
been improved and made available for theories with continuous gauge
groups~\cite{Langfeld:2012ah}. The new method (called LLR method in
the following) features an exponential error suppression and directly
provides access to the partition function. It has been tested for the
$U(1)$, $SU(2)$ and $SU(3)$ gauge theories with large
volumes~\cite{Langfeld:2012ah}.

In this letter, we generalise the LLR-method for the calculation of
the Polyakov line effective potential with an unprecedented precision
due to the exponential error suppression. We will use our numerical
findings to study the properties of $SU(2)$ gauge theory at finite
densities of heavy quarks.
\begin{figure}[t]
\includegraphics[height=6.5cm]{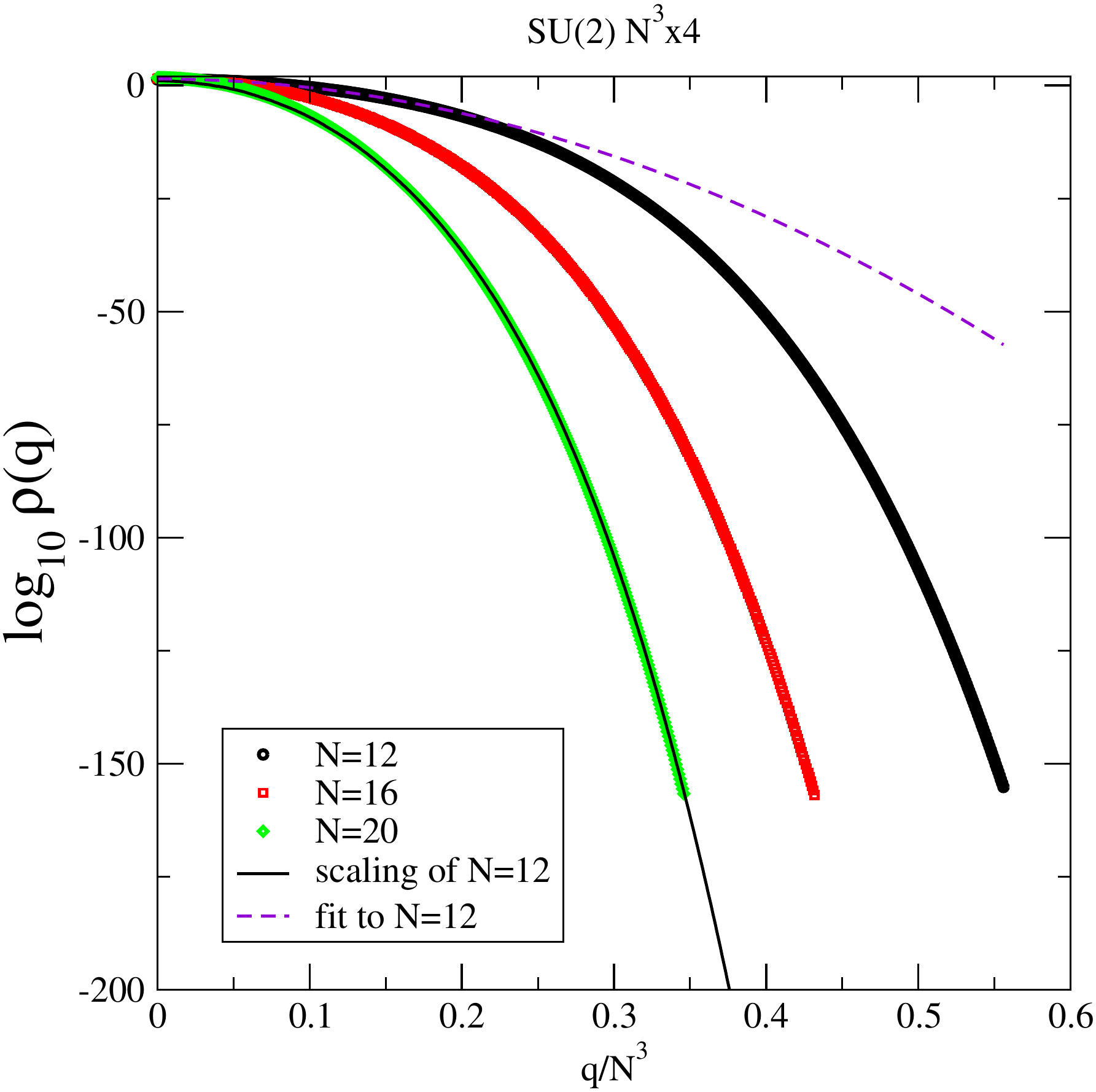} %\hspace{0.5cm}
\caption{\label{fig:0} The Polyakov line probability distribution $\rho (q)$
  (\ref{eq:10}) for a lattice size of $N^3 \times 4$, $N=12,\, 16, \,
  20$ in the confinement phase ($\beta  =2.2 $). 
}
\end{figure}
Our simulations are based upon the Wilson action using a $N \times N_t$
lattice. The gluonic degrees of freedom are represented by 
unitary matrices $U_\mu (x)$ associated with the links of the lattice. 
The Polyakov line $P(\vec{x})$ is the ordered product of the 
time-like links, 
\be 
P(\vec{x}) \; = \; \frac{1}{2} \, \tr \, \prod _{t} \, U_0(\vec{x},t) \; . 
\label{eq:1}
\en The effective potential $V$ is derived by a Legendre
transformation from the generating functional $Z[j]$, 
\bea Z[j] &=&
\int {\cal D}U_\mu \; \exp \left\{ S_g[U] + j \, \sum _{\vec{x}}
  P(\vec{x}) \right\} ,
\label{eq:2} 
\end{eqnarray} 
with 
\begin{eqnarray}
V(q) &=& \frac{T}{\Omega_3} \, \Bigl(  j \, q - \ln Z[j]  \Bigr) \; , 
\qquad  
q = \frac{d \, \ln Z[j] }{dj } \; . 
\label{eq:4} 
\end{eqnarray} 
Here $S_g$ is the Wilson action, and $\Omega_3$ is the spatial volume. 
We also consider a non-vanishing chemical potential $\mu $ 
which couples to heavy and static quarks. In the heavy quark limit  
the quark determinant can be explicitly calculated~\cite{Langfeld:1999ig}. The
complete action can be approximated by 
\begin{figure}[t]
\includegraphics[height=7cm]{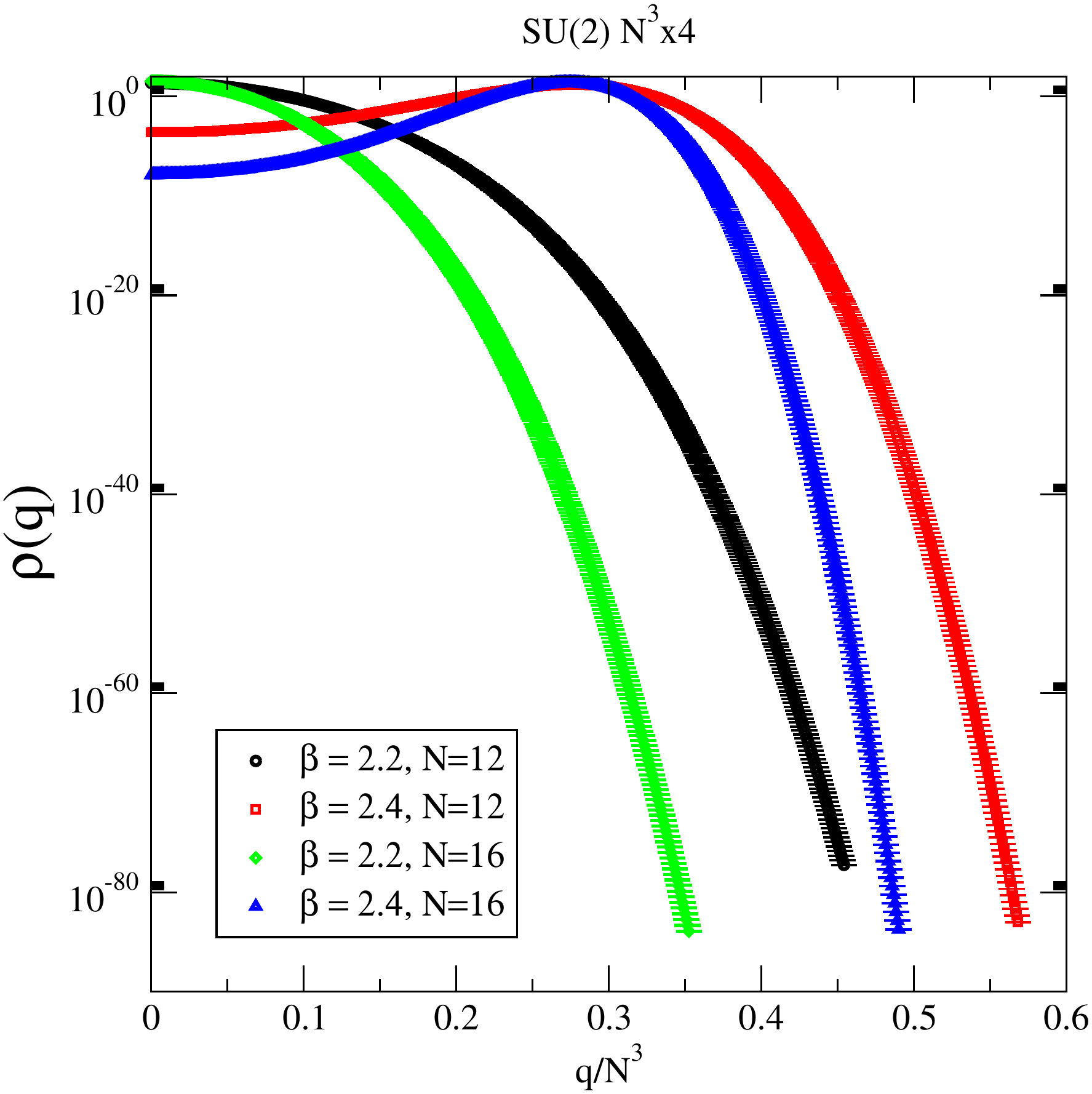} %\hspace{0.5cm}
\caption{\label{fig:1} The Polyakov line probability distribution $\rho (q)$
  (\ref{eq:10}) for a lattice size of $N^3 \times 4$, $N=12,\, 16$ in the
  confinement phase ($\beta  =2.2 $) and in the deconfined phase ($\beta
  =2.4$). 
}
\end{figure}
\bea 
S[U] &=& S_g[U] \; + \; f\, \sum _{\vec{x}} P(\vec{x}) \; , 
\label{eq:5} 
\end{eqnarray} 
with the `relative fugacity' $f$, 
\begin{eqnarray} 
f &=& \frac{ \sqrt{2} }{\pi ^{3/2}} (m T)^{3/2} \, a^3 \, 
\exp \left\{ \frac{\mu -m }{T} \right\} \; . 
\label{eq:6} 
\end{eqnarray}
Here, $m$ is the heavy quark mass, $T$ the temperature and $a$ the lattice 
spacing. Note that the derivation of the result \eq{eq:5} neglects
terms with multiple windings of the Polyakov line around the
torus. This approximation is only justified for chemical potentials
smaller than or close to the mass gap~\cite{Langfeld:1999ig}. As a result, we do not
observe the saturation of the baryonic density for $\mu \gg m$ which
takes place on finite lattices when the  lattice is filled with quarks. 
Here, we consider~\eq{eq:5} as a "weak coupling" extension
of the popular Polyakov loop
model~\cite{Mercado:2011ua,Gattringer:2011gq} 
and take it for granted for the
remainder of the paper. With this assumption, $f$ merely
shifts the external source $j$ in (\ref{eq:2}), and we find  
\be 
V(q;f) \; = \; V(q) \; + \; T \, f \, q / \Omega_3 \; . 
\label{eq:7}
\en 
It is generically difficult to calculate the effective potential $V(q)$
defined in (\ref{eq:4})  by means of Monte-Carlo simulations. The reason for this is that 
the Polyakov line $q$ in (\ref{eq:4}) is an extensive observable, which scales
with the spatial volume: 
$$ 
q \; = \; N^3 \, \langle P (\vec{x}) \rangle (j) \; . 
$$
Hence, the term $\exp\{jq\}$ in the partition function is generically
large, i.e., of order $N^3$, but is cancelled to a good deal by the
``classical term'' in (\ref{eq:4}), i.e., $- j q$, leaving us with a
poor signal-to-noise ratio. 
%One possibility of increasing the
%statistics would be the calculation of the per-site potential
%\cite{SDPS}. 
%
\begin{figure}[t]
\includegraphics[height=7cm]{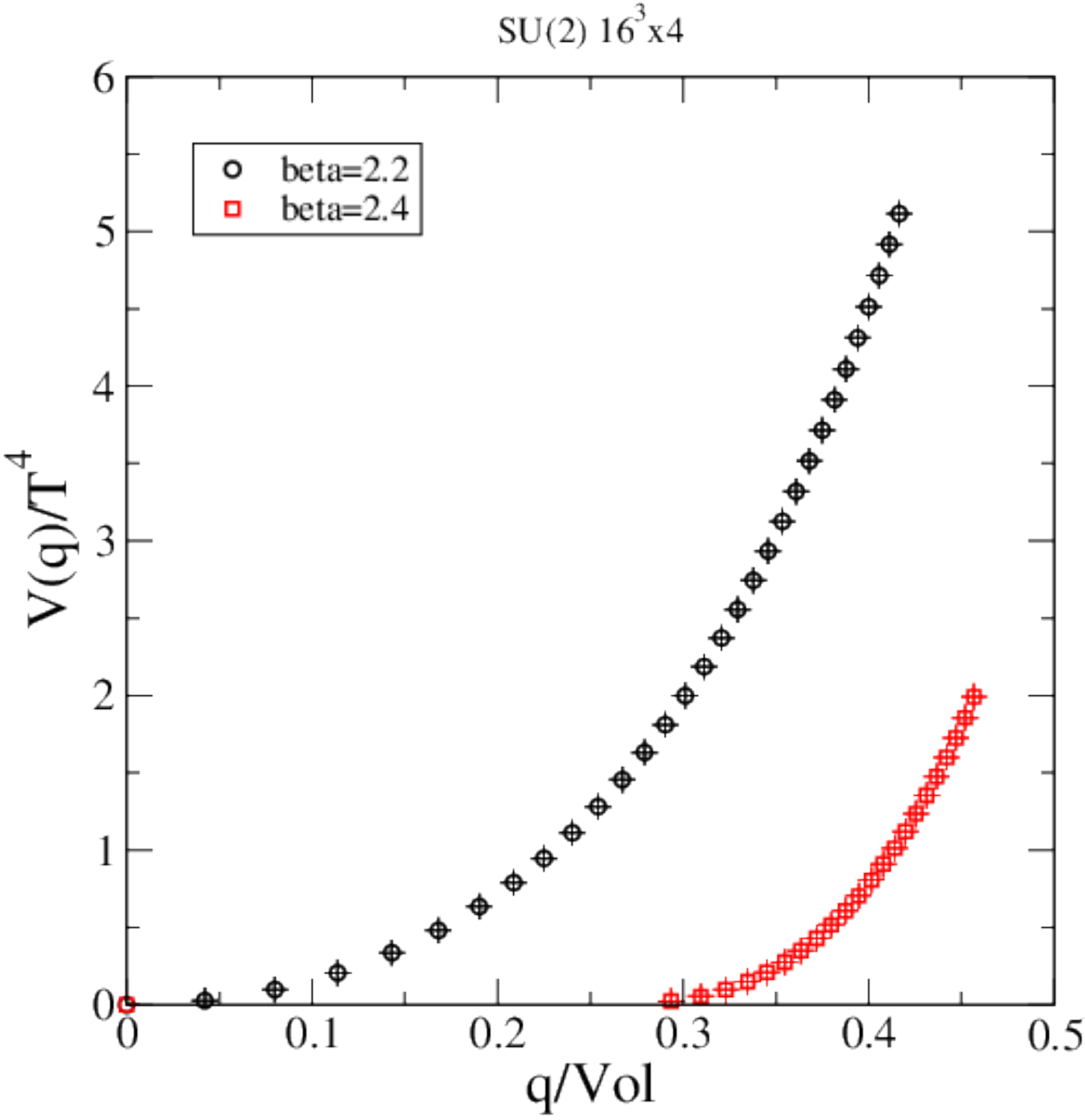} %\hspace{0.5cm}
\caption{\label{fig:2} The Polyakov line effective potential $V (q)$
  (\ref{eq:4}) for a lattice size of $16^3 \times 4$ for $\beta =2.2 $ and
  $\beta =2.4$. 
}
\end{figure}
In the present work we adopt the improved Wang-Landau approach
presented in~\cite{Langfeld:2012ah} for the calculation of the
partition function $Z[j]$ to a very high precision. Central to this
approach is the generalised density of states for the Polyakov line,
\be \rho(q) \; = \; \int {\cal D}U_\mu \; \mathrm{e}^{S_g[U]} \;
\delta \Bigl( q - p[U] \Bigr) \; ,
\label{eq:8}
\en 
where 
$$ 
p[U] \; = \; \sum_{\vec{x}} P(x) \; , 
$$
with $P(x)$ defined in (\ref{eq:1}). If we manage to obtain a high precision 
result for $\rho (q)$, the partition function is recovered by standard
integration, 
\be 
Z[j] \; = \; \int dq \; \rho(q) \; \exp \{ j \, q \} \; . 
\label{eq:9}
\en 
To obtain the density-of-states $\rho $ numerically, we approximate its 
logarithm by a piecewise linear function, 
\be
\rho (q) \; = \; \rho(q_0) \, \exp \Bigl\{ a(q_0) \, (q-q_0) \Bigr\}\,,
\label{eq:10}
\en
for $q_0 \le q < q_0 + \delta q $. We generalise the LLR approach and define
the truncated and re-weighted expectation values of present case by 
\begin{equation}
{\langle \kern-.17em \langle} f(q) {\rangle \kern-.17em \rangle }(a)
= \frac{1}{\cal N} \int_{[q_0,\delta q]} \hskip -0.5cm
{\cal D}U_\mu \, 
\mathrm{e}^{S_g[U]} \, f\Bigl(p[U] \Bigr) \; \mathrm{e}^{-a p[U]}  ,
\label{eq:11} 
\end{equation}
with 
\begin{eqnarray}
{\cal N} &=& \int_{[q_0,\delta q]} {\cal D}U_\mu \; \mathrm{e}^{S_g[U]} \; 
\mathrm{e}^{-a p[U]} \; .
\label{eq:12}
\ena
These expectation values can be calculated by straightforward Monte-Carlo 
simulations. Thereby, the integration only extends over configurations 
$\{U\}$ with $q_0 < p([U]) < q_0 + \delta q$. In practice, this constraint is 
implemented as follows: we start with an `empty vacuum' configuration
 where all links are set to the unit element. We then 
set the time-like links at the time slice $t=1$ according to,  
\begin{equation}
U_0(1,\vec{x})= \cos \theta + i \sin \theta \tau^3, 
\end{equation}
with
\begin{equation}
\cos \theta = \left(q_0+ \frac{\delta q}{2} \right) /N^3 . 
\end{equation}
This generates an `allowed' configuration with $p[U]$ inside the desired 
interval. Subsequent Markov chain updates make sure 
that an update is disregarded (probability zero) if the configurations 
$p[U]$ would fall outside the interval. We then follow the standard
procedures: after thermalisation, we perform``dummy'' sweeps between
measurements. For the determination of $a(q_0)$, we follow the technique
detailed in~\cite{Langfeld:2012ah}. This technique exploits the fact that we find  
$$ 
{\langle \kern-.17em \langle} \Delta q {\rangle \kern-.17em \rangle }(a) = 0 ,
\hbo \delta q = q - \left(q_0+ \frac{\delta q}{2} \right) \,,
$$
if the re-weighting factor $\exp \{-a p[U]\}$ compensates the true
density-of-states. 
This is a non-linear equation which we solve for $a$ using the iteration: 
$$ 
a_{n+1} \; = \; a_n \, + \, \lambda \, \frac{12}{\delta q^2} 
{\langle \kern-.17em \langle} \Delta q {\rangle \kern-.17em \rangle }(a_n). 
$$
Thereby, $\lambda <1 $ is an under-relaxation factor which we found useful to
control the amount of noise in during the iteration. The precise value of 
$\lambda $ is not critical, and we worked with $\lambda =0.25$ most of the
time. We repeat these steps for any $q_0$ out of the range of interest and
reconstruct the density-of-states $\rho (q)$ (\ref{eq:10}), i.e., the 
Polyakov line probability distribution. 
 
We have studied the Polyakov line distribution function  $\rho (q)$
(\ref{eq:10}) for several lattice sizes $N^3 \times 4$ for $\beta =2.2$ and for 
$\beta =2.4$. For this temporal extent, the deconfinement phase
transition occurs for $\beta \approx 2.3$ which leaves us with $\beta =2.2$ in
the  confinement phase and for $\beta =2.4$ well in the high temperature
deconfined phase. Let us firstly consider the confinement phase. 
In order to check our new numerical method, we have also calculated $\rho (q)$ 
by histogramming the Polyakov line obtained by a standard heat-bath
simulation. In the small $q$ range, $q<0.05$, where a reasonable of amount of
statistics can be obtained by the standard method, we find a good
agreement. We stress that our new method can easily produce results for $q$ as
large as $0.4$ and provides access to the probability distribution over more 
than hundred orders of magnitude with good precision. 
The (logarithms base $10$ of the) probability distributions $\rho (q)$ are 
shown in Figure~\ref{fig:0} for several spatial values. The error bars have been
obtained with the bootstrap method and are well with the plotting symbols. 
As expected, the probability distributions centre around the trivial value 
$q=0$. The distributions get narrower with increasing spatial volume. In fact,
one expects that $\mathrm{log}_{10} \rho (q)$ scales with number of degrees 
of freedom and therefore with the volume $\Omega_3$. To put this to the test, we have
rescaled the $N=12$ results to match with the $N=20$ distribution. After a
proper normalisation of the distribution, the result is shown in
Figure~\ref{fig:0}, solid line. We find an almost perfect match. We also point
out that the distribution $\rho (q)$ of the Polyakov line is far from
Gaussian. To this aim, we have fitted the low $q$ range of $\mathrm{log}_{10}
\rho (q)$ at $N=12$ to a parabola. The result of this fit (dashed line in
Figure~\ref{fig:0}) shows large deviations from the quadratic behaviour even
at moderate values of $q$. 
\begin{figure}[t]
\includegraphics[height=6.5cm]{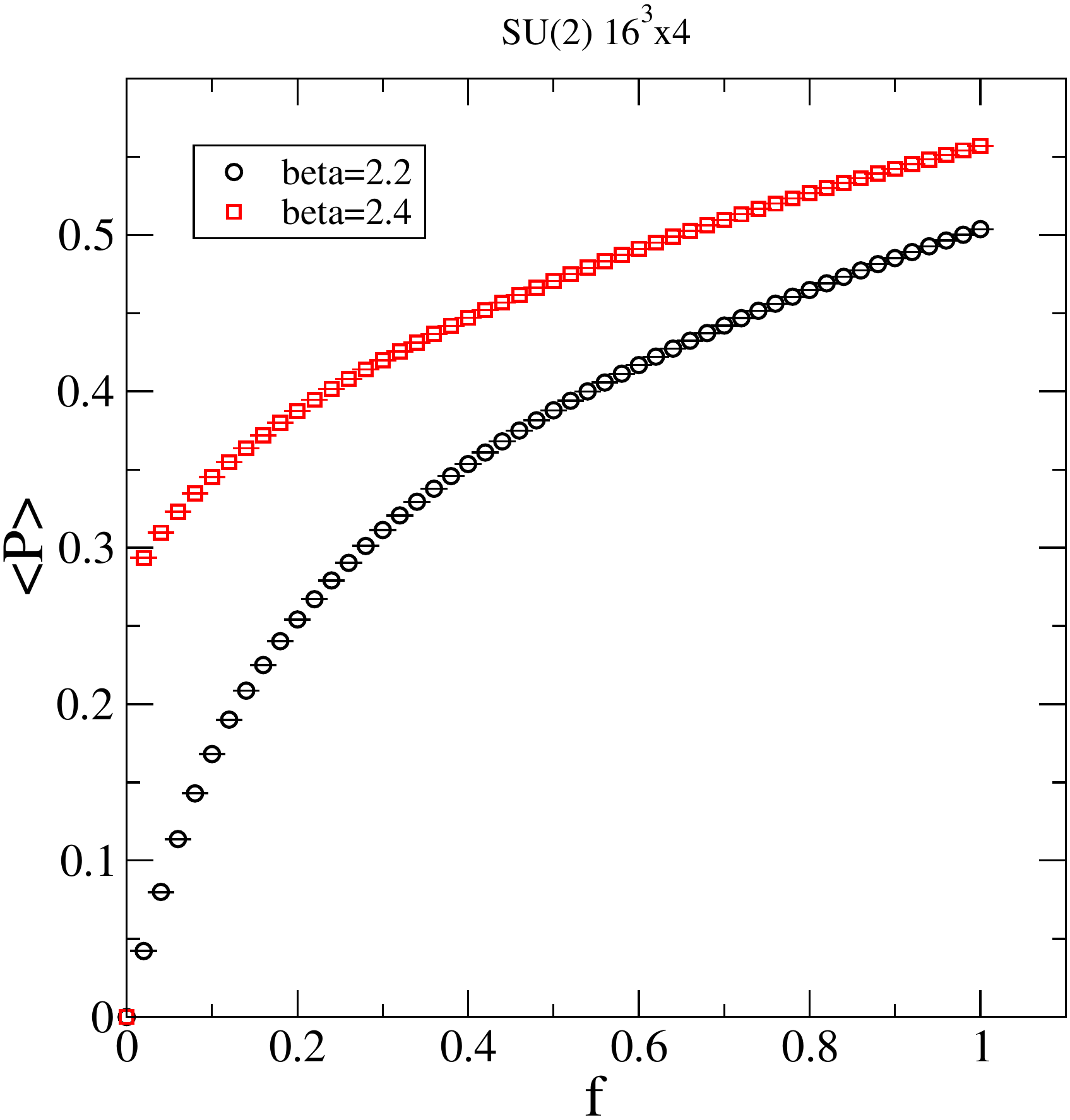} %\hspace{0.5cm}
\caption{\label{fig:3} The Polyakov line expectation value  $P$ for a $16^3
  \times   4$ lattice, $\beta =2.2 $ and   $\beta =2.4$. 
}
\end{figure}

The probability distributions of the confined and the high-temperature phase 
are compared in Figure~\ref{fig:1}.  In the latter phase ($\beta=2.4$), the
most likely value for $q$ is different from zero indicating the spontaneous
breakdown of the centre symmetry. Again the distribution functions get
narrower with increasing spatial volume leaving us with a decreasing value 
for $\rho $ at $q=0$. One expects that $\rho (q=0)$ vanishes in the infinite 
volume limit realising the spontaneous symmetry breakdown. A more systematic
volume study of this effect will be presented elsewhere. 

It turns out that the precision of the data is good enough to attempt the
Legendre transformation and to directly obtain the Polyakov line effective
potential. To this aim, we choose $j$ as an implicit parameter and obtain the  
partition function by means of (\ref{eq:9}) and $q$ by 
$$ 
q(j) \; = \; \frac{ \int dq^\prime \; \rho(q^\prime) \; q^\prime \; \exp \{ j \,
  q^\prime \} }{ \int dq^\prime \; \rho(q^\prime) \;  \exp \{ j \,
  q^\prime \} }.
$$
Substituting both into (\ref{eq:4}) allows us to obtain the potential in
implicit form. Error bars are obtained by the bootstrap method. The related 
result is shown in Figure~\ref{fig:2}. In both cases, i.e., 
$\beta=2.2$ and $\beta=2.4$, the potential is convex. For the deconfinement
case, the potential is flat until $q \approx 0.3$ indicating the spontaneous
breakdown of centre symmetry. The absolute values of the potential in
units of the temperature are quite high when compared to the results
from ab initio continuum methods. We stress that only the
unrenormalised potential as function of the unrenormalised Polyakov
line is shown in Figure~\ref{fig:2}. To prepare the grounds for a
comparison, simulations with more values of $\beta $ and several
lattice sizes are need to renormalise and extrapolate to the continuum
limit. This is left to future work. 

Now we apply our approach to two-colour QCD at finite densities of heavy
quarks. The baryon number $B$ can be retrieved from the partition function 
by 
\be 
B \; = \; T \, \frac{ d \ln Z }{d \mu } \; = \; f \, \langle p[U] \rangle \; , 
\label{eq:15}
\en 
where we have used (\ref{eq:5}) and (\ref{eq:6}). We will use the heavy quark 
mass $m$ as the fundamental scale to present our results: 
We define the baryon density (in units of $m$) as  
\bea 
\rho _B / m^3 &:=& B/( V m^3) 
\nonumber \\ 
&=& \sqrt{2} \, \left( \frac{T}{m \pi} \right)^{3/2}  
\mathrm{e}^{ (\mu -m )/T } \; \frac{1}{N^3} \langle p[U] \rangle 
\; . 
\label{eq:16}
\ena 
In the heavy quark limit (and within the remit of the approximations leading
to (\ref{eq:5}), $\rho _B $ is proportional the Polyakov line 
expectation value. Our high precision results for $\langle P \rangle  =
\langle p[U] \rangle  /N^3$ are shown in Figure~\ref{fig:3}
for the phases below and above the deconfinement transition. While $\langle P
\rangle$ goes to zero for vanishing fugacity $f$ in the confinement phase,
$\langle P \rangle$  remains finite in the same limit for the high
temperature phase. The reason 
for this is that any small non-vanishing value for $f$ triggers a sizeable
response for $\langle P \rangle $ because of the
spontaneously broken centre symmetry in the high temperature phase. 

\begin{figure}[t]
\includegraphics[height=6.5cm]{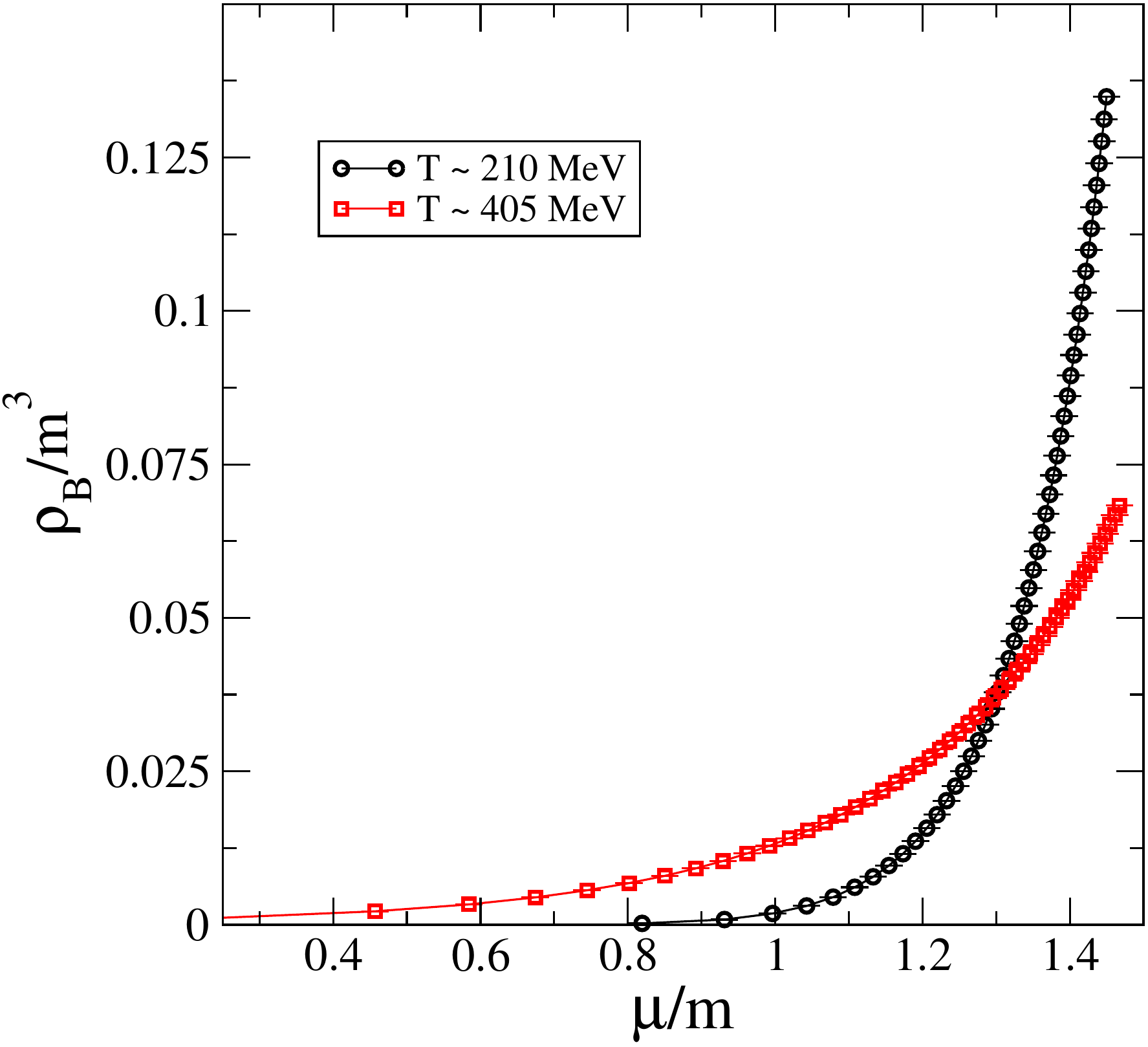} %\hspace{0.5cm}
\caption{\label{fig:4} The baryon density $\rho _B$ for "charmonium" 
as a function of the chemical potential $\mu $ in units of the charmonium mass
$m$ in the confinement phase $(T\approx 210 \mathrm{MeV})$ and 
in the gluon plasma phase $(T\approx 405 \mathrm{MeV})$. 
}
\end{figure}
Let us now investigate the response of the baryon density to the chemical
potential for the two-colour analogue of charmonium with $m = 1.29 \, $GeV. For vanishing 
temperature, the fugacity $f$ vanishes as well for chemical potentials 
smaller than the mass gap, i.e., $\mu < m$. Thus, the onset value coincides
with the mass $m$, and a Silver Blaze problem~\cite{Cohen:2003kd} is
absent in our approach.  In the following, we consider two temperatures well above and 
well below the confinement-deconfinement temperature. 
The temperature is generically given  
by 
$$ 
T \; = \; \frac{1}{N_t a} \; = \; \frac{\sqrt{\sigma }}{ N_t 
\sqrt{\sigma a^2} } \; , 
$$
where the lattice spacing is measured in units of the string tension. 
We use $\sigma a^2 (\beta =2.2) = 0.28(1)$ and $\sigma a^2 (\beta =2.4) = 
0.0738(5)$, see e.g.~\cite{Langfeld:1999ig}. With the standard value
$\sigma = (440 \mathrm{MeV})^2 $  and for a $16^3 \times 4$, $N_t=4$
lattice, we are left with temperatures $T(\beta=2.2) = 210 \,
$MeV, well below transition temperature $T_c \approx 300 \,$
MeV, and   $T(\beta=2.4) = 405 \, $MeV in the deconfinement phase. Our
findings are summarised in Figure~\ref{fig:4}: In the confinement phase and for chemical
potentials below the onset value, the response of the density to $\mu $ is
much weaker than in the gluon plasma phase. The physics behind this
observation is that in the confinement phase temperature effects need to
excite a baryon whereas in the plasma phase single quark excitations are
allowed.

Let us now discuss the existence of a density driven phase transition
in heavy quark SU(2) Yang-Mills theory. Within the remit of the
approximations leading to the effective theory \eq{eq:5}, we do not
expect that a phase transition exists: Since the fugacity $f$ directly
couples to the Polyakov line, any small value implies that Polyakov 
line potential flattens at large distances $\vert \vec{x} \vert$, 
\begin{equation}
  \langle P(\vec{x}) P(0) \rangle \to \langle P \rangle ^2 (f) \not= 0 \, . 
\end{equation}
This property relates to the breaking of the colour electric
flux tube: the colour electric flux tube, which connects a quark and a
anti-quark test charge, can break and attach to one of the static background
charges. Hence, deconfinement sets in at the onset value $\mu = m$. 
The situation is similar to the $O(2)$ quantum model for which superfluidity 
occurs at the critical chemical potential~\cite{Langfeld:2013kno}. We point out 
that lattice simulations and effective model computations of dense two-colour QCD 
found a more rich state of matter for chemical potentials close to the onset
value, for recent results see e.g. \cite{Cotter:2012mb,Boz:2013rca,Strodthoff:2013cua}. 
This is (i) due to the finite quark masses in Q${}_2$D and (ii) due to the neglect of 
higher Polyakov windings in the
built-up to \eq{eq:5}. These contributions are potentially important
to describe physics such as the transition from BEC to BCS. \\[-2ex]

%\vskip 1mm
%\par\bigskip
\noindent {\bf Acknowledgments:} 
We thank  Antonio Rago for helpful discussions. 
This work is supported by STFC under the DiRAC framework, by the Helmholtz Alliance
HA216/EMMI and by ERC- AdG-290623. We are
grateful for the support from the HPCC Plymouth, where the numerical
computations have been carried out.

\bibliography{polyak_eff}{}

\end{document}